# MEM as an Extension of Recurrent Processing Theory: Toward a Motivational Account of Consciousness

Wiesław L. Galus & Janusz A. Starzyk

*Secondary perception, semblions, embodiment, emotions, memory, competition among representations, and the biophysical conditions of phenomenal consciousness*

**Abstract**

Victor Lamme's Recurrent Processing Theory (RPT) proposes that the feedforward sweep (FFS) can extract and categorize visual features unconsciously, while conscious vision emerges through recurrent processing and feedback interactions across representational levels. The Motivated Emotional Mind (MEM) model, developed by Wiesław L. Galus and Janusz A. Starzyk, retains this core principle but embeds it in an embodied architecture integrating multilayer networks, semi-hierarchical semblions, receptors, effectors, allostasis, emotional valuation, and memory-based sensory reconstruction.

This article examines the substantive but limited relationship between RPT and MEM. While RPT explains the role of recurrence in perceptual organization and the transition to conscious vision, MEM addresses the content of recurrent processing, its grounding in receptor activity and bodily states, its interaction with memory and action, and the possible role of reconstructed sensory representations in phenomenal experience. Particular emphasis is placed on secondary perception: the top-down reconstruction of sensory and interoceptive maps through which memories, mental imagery, dreams, hallucinations, feelings, and observed scenarios acquire experiential form. MEM thus associates phenomenality not merely with recurrent activity or reportability, but with the reconstruction of modality-specific, receptor-grounded sensory and bodily content.

The article also presents the mathematical model of MEM, formalizing re-entrant processing, representational competition, allostasis, affect, curiosity, procedural gaps, and action selection. We conclude that MEM can be viewed as an embodied and motivational extension of RPT. However, its distinctive claims, particularly the sufficiency of secondary perception for phenomenality and the proposed microstructure of semblions, remain hypotheses requiring causal testing.

## 1. Introduction: Why Comparing RPT and MEM Matters

A central problem in contemporary consciousness research is to identify the neural processes that distinguish conscious experience from processing that can occur without awareness. RPT has played an important role in the neuroscience of consciousness by shifting the discussion from the question of where consciousness is located to the question of how information is processed. Lamme and Roelfsema (2000) contrasted the rapid feedforward sweep (FFS) with subsequent recurrent processing involving feedback and horizontal interactions. This distinction helped explain why the brain can detect features, categorize stimuli, and even prepare responses although the subject does not report seeing them. Consciousness would therefore arise not merely from representations appearing in progressively higher areas, but from renewed, multidirectional interactions between representational levels (Lamme & Roelfsema, 2000; Lamme, 2006, 2010, 2018).

The significance of RPT also lies in its distinction between phenomenal consciousness and access, reporting, and attention. In Lamme's account, local recurrent processing may suffice for visual experience before its content becomes globally available for reporting, working memory, or task control. Attention may influence the selection and amplification of content without necessarily being a condition of experience itself. This position has made RPT a major competitor to global access theories and higher-order theories.

At the same time, its strength is closely tied to its scope: RPT originated as a theory of conscious vision, based primarily on studies of figure - ground organization, masking, attention, and visual cortical activity. This does not mean that RPT excludes the influence of emotions, bodily states, or other neural systems. On the contrary, Lamme allows for their involvement in shaping the conscious percept. However, these elements do not constitute a complete architecture of an emotional, embodied, and acting mind within RPT.

MEM emerges from the same insight into the importance of feedback processing but gives it a different scope and ambition. Galus and Starzyk's Reductive Model of the Conscious Mind (2021) presented an architecture linking the brain, receptors, body, and effectors. Subsequent publications developed distinct processes corresponding to different aspects of consciousness, the concept of mind - brain identity, motivations, a perceptual account of emotions, phenomenality as a result of secondary perception, semi - hierarchical semblions, and dynamic image schemas and thinking processes (Galus, 2018, 2024a, 2024b, 2025a, 2025b, 2026; Galus & Starzyk, 2021). Consequently, MEM broadens the question: it concerns not only when perceptual processing becomes conscious, but also how an embodied system can recognize perceived objects and generate feelings, mental images, action priorities, and deliberative processes.

RPT and MEM share the conviction that unidirectional feedforward processing does not exhaust the mechanism of conscious experience. MEM accepts this premise but proposes an additional level of explanation: the content of recurrence is to be linked to previously formed sensory representations, the organism's state, memory, affective value, and possible actions. From the MEM perspective, RPT stops too early: it identifies recurrent processing but does not sufficiently develop the mechanism of secondary perception, the biological grounding of emotions, multilayer associative memory, or the transition from perception to deliberation and action.

The thesis of this article is as follows: MEM specifies the sources and functions of the content involved in recurrence and embeds it within an architecture encompassing the body, memory, emotions, motivation, and action. A substantial part of the MEM architecture can be interpreted as elaborating a question left open by RPT: what exactly is reconstructed during recurrent processing, and why might such reconstruction be associated with phenomenal experience?

The remainder of the article first presents the basic assumptions of RPT in greater detail. It then discusses the MEM architecture, the concept of a semblion, competition between *bottom-up* and *top-down* processes, secondary perception, and the roles of interoception, allostasis, emotions, and action. The final sections compare the two models and examine which MEM claims can be treated as elaborations consistent with RPT and which constitute stronger, distinct hypotheses requiring their own empirical tests.

## 2. RPT: From the Feedforward Sweep to Conscious Perceptual Organization

### 2.1. Two Modes of Vision

During the FFS, activation propagates from the retina through successive levels of the visual system to higher areas. Within approximately the first 100–150 ms, orientation, colour, motion, complex features, and object categories can be extracted. Lamme emphasizes that many operations hardwired into feedforward connections do not require consciousness. In his 2020 paper, he explicitly states that scene features and categories are extracted during the feedforward sweep “regardless of consciousness” (Lamme, 2020, p. 4). This means that merely detecting or classifying a stimulus cannot be equated with conscious vision.

Recurrent processing begins when representations in higher and lower areas start to interact. Feedback and horizontal connections enable grouping, figure–ground segregation, contour completion, contextual processing, and the stabilization of an extensive scene representation. Figure–ground studies have shown that the initial V1 response to local features may remain present even when the later modulation associated with perceptual organization disappears under conditions of reduced awareness (Lamme et al., 1998; Lamme, 2020). These findings suggest that subsequent recurrent processing is more closely associated with the conscious percept than the FFS alone.

In his 2015 paper, Lamme summarizes the functional properties of this stage as STERP, SUPER, and LEARN. STERP denotes processing extended in space and time; SUPER denotes a composite representation formed by integrating information distributed across multiple modules; and LEARN denotes the plasticity and learning through which conscious solutions can become automated (Lamme, 2015). Conscious vision does not depend on a simple feedback impulse: what matters is multistage, extensive, integrative dynamics.

From the MEM perspective, this is particularly important. If recurrence is a genuine mechanism of conscious processing, we need to know which representations participate in the loop, where they come from, and how their interaction give rise to a particular percept. This question is central to the MEM concept of secondary perception.

### 2.2. Phenomenality, Access, and Attention

Lamme distinguishes at least four states of vision: complete stimulus invisibility, subjective invisibility, conscious but unattended vision, and vision available to attention and reporting (Lamme, 2020). RPT therefore allows for phenomenal consciousness without full access. Attention may amplify, select, and make content available, but the theory does not regard it as the sole source of that content.

RPT faces an inferential problem: if experience is independent of reporting, how can its presence be established empirically? Critics point out that local recurrence may also accompany unconscious processing and that some supposedly feedforward processes may suffice for more complex behaviour than previously assumed (Tsuchiya & van Boxtel, 2010). Cohen and Dennett (2011), in turn, argued that separating consciousness too sharply from function risks rendering it unfalsifiable. Doerig et al. (2021) formulated the more general "unfolding" objection: different causal architectures can generate identical input–output relations, so a theory must provide criteria for distinguishing them empirically.

This criticism does not invalidate the findings concerning recurrence. It does, however, require three claims to be distinguished: that recurrence correlates with conscious vision, that it is causally necessary for it, and that it is sufficient for phenomenality. The first is strongly supported by visual perception research. The second requires selective perturbations of feedback pathways while preserving the FFS. The third is the most difficult, because the mere presence of feedback does not yet explain why a process has a felt quality. This is precisely where the MEM proposal becomes relevant. MEM does not assume that all recurrence is conscious. It proposes an additional condition: recurrence associated with phenomenality should produce the appropriate reactivation of receptor-grounded sensory and interoceptive maps.

### 2.3. What RPT Claims—and What Should Not Be Attributed to It

RPT is not a naively disembodied theory. Lamme (2020) acknowledges that experiences are shaped by emotions and bodily states and that the integration of visual information with emotional, motor, and interoceptive information may depend on recurrent interactions. The difference between RPT and MEM primarily concerns the status of these elements within the explanatory architecture. In RPT, they are possible extensions of the conscious percept, whereas in MEM they become constitutive parts of the subject's architecture: bodily state, interoception, memory, affect, motivation, and action influence competition among representations and determine which content undergoes recurrent reconstruction.

Unlike MEM, RPT does not provide a developed theory of long-term semantic and episodic memory, allostatic weighting of representations, motivated action selection, or deliberation across multiple cycles. Nevertheless, focusing on such omission would be methodologically unfair since RPT theory has a narrower scope. An appropriate comparison should therefore ask whether MEM adds mechanisms that generate new, discriminating predictions (Barrett & Simmons 2015). In particular, it is necessary to examine whether recurrence alone is sufficient for phenomenality or whether a specific source and mode of content reconstruction are also required.

Thus, RPT provides a mechanism of recurrent organization, whereas MEM seeks to specify the content, grounding, and functional context of that organization. This difference is the basis for the further comparison of the two models.

**3. The Development of MEM and Its Reductive Ambition**

The genealogy of MEM comprises several stages. Early work on the architecture of consciousness and the concept of “word semblions” developed the idea of multilayer representations and associative transitions from perception to concepts and language (Galus, 2015a, 2015b, 2015c, 2018). The book Reductive Model of the Conscious Mind constructed an integrated architecture linking the brain with receptors, the body, and effectors and introduced motivated learning as the basis of autonomous activity (Galus & Starzyk, 2021). Subsequent publications distinguished biophysical processes corresponding to different aspects of consciousness and developed the identity of mental and brain processes, a perceptual theory of emotions, and a mechanism of phenomenality based on feedback activation of lower-level sensory fields (Galus, 2024a, 2025a, 2025b).

The 2026 article systematizes MEM as a model of thinking in its cognitive and phenomenal aspects. It encompasses recognition, memory, planning, inference, thinking, dynamic semblions, image schemas, and relations between the brain, receptors, and effectors (Galus, 2026). The article emphasizes the importance of “dynamic interactions between the brain, sensory receptors, and effectors”, integrated through the system’s embodiment (Galus, 2026). In this way, MEM changes the explanandum: it asks not only when a percept is conscious, but also how an embodied network can generate recognition, feeling, priority, imagery, and action.

The reductive ambition of MEM is to equip an organism with a specific organization: multilevel neuronal–glial fields, plastic connections, receptors, regulatory systems, effectors, and competing activation pathways whose operation is driven by the organism’s needs and motivations (Graham et al., 2014; Raif & Starzyk, 2011; Starzyk, 2008; Starzyk & Graham, 2017; Starzyk et al., 2012; 2017).

On this account, selection, categorization, association, and the initiation of action are expected to emerge as consequences of system dynamics. These processes do not occur with logical necessity in every network but depend conditionally on its architecture, dynamic parameters, learning history, and the organism’s current allostatic state, which is related to the degree to which its needs are satisfied.

The reductive ambition of MEM should therefore be understood as an attempt to show that higher cognitive and phenomenal functions can be explained through a lower level of physical organization and its dynamics. This is relevant to the comparison with RPT: RPT also seeks the mechanism of consciousness in neural dynamics, but focuses primarily on the organization of the conscious percept. MEM seeks to extend the reductive explanation to memory, emotions, motivation, imagery, and action.

MEM simultaneously invokes several levels of biological organization. At the molecular and synaptic level, it refers to changes in connection strength and synaptic structure, dendritic spines, long-term potentiation, and the involvement of astrocytes in the tripartite synapse (Matsuzaki et al., 2004; Perea et al., 2009; Harris, 2020). At the mesoscopic level, it postulates semblions—distributed, multilayer representational coalitions. At the systems level, it links sensory, memory, affective, motor, and regulatory networks.

This multilevel approach connects mechanisms of plasticity and neural processing with functional systems. The neurobiological literature supports constituent components of MEM architecture but does not yet provide direct evidence that the semblion organization proposed is the actual implementation of consciousness.

## 4. Semblions: Semi-Hierarchy, Compression, and a Model of Reality

### 4.1. From Sensory Maps to Concepts

A central concept in the MEM architecture is the semblion—a semi-hierarchical neural structure spanning multiple layers and modalities. Its lower components are linked to receptor and sensory configurations; its higher components are linked to objects, concepts, contexts, emotions, and **action programs**.

"Semi-hierarchy", or heterohierarchy, emphasizes that connections do not run exclusively upwards: feedback, lateral, and cross-modal pathways are also present. Such organization is consistent with the distributed hierarchy in the primate cortex (Felleman & Van Essen, 1991), but the concept of a semblion and its organization constitute a specific MEM hypothesis rather than a standard anatomical term.

Information compression is an important property of this architecture. Compression results from convergence: many elementary lower-level fields can jointly constitute a more abstract pattern, and similar configurations are subsumed under a category. This mechanism resembles the principles of sparse and efficient coding, in which a representation preserves regularities relevant to recognition instead of storing each episode as an independent copy (Barlow, 1961; Olshausen & Field, 1996).

Findings concerning concept cells in the medial temporal lobe demonstrate a high degree of invariance and selectivity in representations of people or objects. This does not, however, mean that a single cell stores an entire concept (Quiroga et al., 2005; Quiroga, 2019; Galus, 2018; 2021). These findings suggest more abstract, distributed representations that, in MEM, are organized into semblion coalitions.

Compression has three main benefits in MEM. First, it increases memory capacity by allowing multiple representations to share features. Second, it supports generalization so new objects can activate patterns partially similar to those already stored. Third, it strengthens associations by linking information from different modalities and episodes. The model of reality is therefore not a collection of images but a network connecting objects, their properties, typical consequences, biological value, and possible actions.

Winner-takes-all competition, Top-K selection, or soft selection help identify the most relevant semblions for the current input and context. In neurobiology, lateral inhibition, rhythmic coordination, excitation–inhibition balance, and basal ganglia circuits may perform analogous functions. MEM need not require a single winner; it requires a dominant group of representations that can guide further reconstruction and action.

#### 4.2. Distributed Memory and Large-Scale Associations

In MEM, memory is not separate from perception. Lasting changes in connections and excitability create engrams within the same structures involved in recognition and ongoing processing. Different types of memory preserve different relationships, such as object properties, event sequences, motor skills, or temporary working content. This view is related to the convergence and divergence of recognition and memory (Meyer & Damasio, 2009), but MEM organizes these processes through semblions.

Large-scale multimodal associations link visual appearance, sound, touch, smell, internal organ states, and the consequences of action. For example, if a person has repeatedly caused fear, recognizing their face may activate not only visual memories but also bodily tension and autonomic responses. These affective associations influence which semblions become dominant. Thus, memory in MEM does more than merely support perception: it gives it biological significance.

This extends RPT, which explains how recurrent processing binds features into a figure–ground percept. MEM also explains why certain semblions dominate: they match not only sensory input but also existing memory patterns, current needs, emotional values, and available action programmes. In this way, MEM extends perceptual organization to include semantic and affective selection.

The difference can therefore be described as a transition from perceptual organization to the semantic and affective selection of representations. This is an important part of the argument for treating MEM as an extension of RPT, because it does not just add further cognitive functions; it assigns recurrent processing a broader role in organizing the entire embodied system.

### 5. Direct Perception and Bottom-Up/Top-Down Competition

In MEM, “direct perception” denotes a process distinct from Gibson’s ecological theory. It begins with the activation of exteroceptors or interoceptors, proceeds through lower-level sensory fields and the FFS, and then matches the configuration to memory patterns in higher layers (Galus, 2024a, footnote 1). Recognition is not a simple mapping of the stimulus. Even at this stage, the system selects among alternative memory configuration under the influence of current sensory data, lateral associations, context, prior experience, and active higher-level representations.

Direct perception, understood in this way, is therefore not a purely bottom-up process. The bottom-up signal provides input, but its interpretation depends on the representations available in memory that best fit the current context. In MEM, higher layers can use feedback to amplify selected components of a representation and weaken competing configurations. The resulting percept emerges from dynamic competition between receptor-driven activation and activity arising from previous learning and the system's current state. This leads to the completion and correction of perceived objects, providing a broader context for both content and emotional state.

Until activation in lower fields is processed in higher layers, the system has access to the object's principal features but not to its full recognition or associated meaning. In higher layers, the current configuration can be linked to established traces of earlier interoceptive activation: pain, tension, satiety, breathlessness, warmth, changes in heart rate, and more indirect indicators of the hormonal or metabolic environment. Some of these states have no single, easily localized receptor; in such cases, MEM refers to a functional "sensory complex" integrating multiple signals (Galus, 2026, footnote 16).

What matters most is the dynamic relationship between two streams. Bottom-up activation is anchored in the current environment and bodily state. Top-down activation originates in active memory representations, mental images, expectations, plans, and moods. The two streams meet in lower and intermediate layers, reinforce or suppress one another, and their relative strength determines whether the directly perceived scene or internally generated content predominates.

The example of a nightclub illustrates this mechanism. Loud sound, rhythmic stimulation, lights, and intense social cues generate bottom-up input of high amplitude and significance. Such a stream can displace subtle memory reconstructions and weak interoceptive nuances, making quiet recollection or the analysis of subtle feelings difficult. This does not imply a complete cessation of thinking, but a change of the competition thresholds. In MEM, this interaction is represented at a higher cognitive level in the functional block responsible for attention switching. The stream of sensory perception directly competes with memory signals for attention.

The opposite situation occurs during deep absorption in thought. Images pass before the eyes and sounds reach the ears, yet an active top-down coalition keeps hold of a memory, a problem, or an imagined scenario, provided it dominates the semblions activated by sensory signals. The FFS may continue to process some stimuli, but they never gain enough strength to dominate the highest layers or influence reporting and action. This explains the everyday phenomenon of inattention and also suggests a continuum leading towards vivid imagery, illusions, and hallucinations when reconstruction predominates over current receptor input.

RPT distinguishes attention from consciousness and explains how percepts compete. MEM further considers how external sensory input competes with content from embodied memory. This competition depends on several factors, including stimulus strength, similarity to stored representations, emotional value, current allostatic needs, associative priming, and action priorities. It can also be influenced by mental saccades — the spontaneous shift of attention initiated by a randomly activated semblion (Starzyk, 2011). Attention is therefore not a separate spotlight but an outcome of competition among activations within a hierarchical network.

## 6. Recurrence as Correction Through Memory and Subconscious Processes

In both RPT and MEM, feedback modifies the representation formed during the FFS. However, the theories differ in how they explain this modification. In RPT, feedback and horizontal interactions primarily implement the percept through grouping, figure–ground organization, completion, and integration. In MEM, active higher-order structures also activate patterns from memory and subconscious processes. The top-down signal is therefore do more than enhance parts of a percept, but they help reconstruct expected sensory patterns and associated bodily states.

This mechanism allows the brain to interpret incomplete, noisy, or ambiguous input. For example, when only part of a contour is visible, the system selects the semblion that best matches the available features and context and then activates the related lower-level components. The resulting percept is richer than the original sensory signal.

However, the same process can also produce illusions when a strong memory pattern imposes an interpretation inconsistent with the stimulus. Thus, the mechanism responsible for correcting or completing an incomplete percept can also lead to erroneous perception.

In the mathematical model of MEM, the relative contributions of bottom-up and top-down processing are described by a reconstruction weighting parameter. When this weight is low and the receptor signal is strong, direct perception predominates. As top-down feedback becomes stronger, perception may shift towards imagery and hallucination (Galus & Starzyk, 2026).

Here, “subconscious processes” denote not a mysterious store but established weights, thresholds, and associations that shape the propagation of activation without a separate report. Preferences, previous experiences, and habits influence semblion matching and access to higher layers. MEM thereby links the unconscious determination of the FFS with the possibility of conscious reflection: the subject does not choose the initial associations, but repeated recurrent cycles can simulate possible consequences and compare alternatives, while the gradual modification of representations through learning can eventually transform established patterns. Recurrence thus becomes not only a mechanism for correcting perception but also part of the process of constructing and updating an internal model of reality.

Neuroimaging shows partial overlap between representations involved in perception and imagery, as well as the presence of modality-specific content in early sensory cortical areas: sound or imagery content can be decoded from early visual cortex, and images maintained in working memory are reconstructed in visual cortex (Harrison & Tong, 2009; Vetter et al., 2014; Xie et al., 2020). These findings do not demonstrate that reactivation is identical to phenomenality. They do, however, provide an empirical basis for asking whether the reactivation of lower-level maps can play a causal role in the emergence of conscious experience.

RPT shows that the recurrent organization of representations plays a crucial role in perception. MEM formulates a stronger hypothesis: higher-level activity can reactivate receptor-grounded lower-level representations and subsequently participate in forming content available as experience.

## 7. Secondary Perception as the Core of Phenomenal Consciousness in MEM

### 7.1. What "Secondary" Means

The distinction between direct and secondary perception can therefore be stated as follows. Direct perception begins with current receptor activation and proceeds through feedforward processing to the formation of a higher-level representation. Secondary perception occurs when an active higher-order representation feeds back to activate lower-level sensory or interoceptive fields, reconstructing a configuration learned during earlier receptor-mediated perception. Recalling a face is then not limited to an abstract label for a person; it may reactivate shape, colour, voice, movement, and the affective state associated with that person. Imagining a melody can activate patterns in the auditory cortex despite the absence of an acoustic wave. Dreams and hallucinations are cases in which top-down content acquires a substantial advantage over external bottom-up stimulation.

The central claim of MEM is that phenomenal consciousness is not merely a report. As stated in the article developing the model, "Phenomenal consciousness is not just a form of report … It is a sensation, not a description" (Galus, 2026, Section 7). A report can describe the content, but its qualitative character is proposed to arise from the activation of the same maps, or functionally continuous maps, maps that were originally formed through receptor activity. Secondary perception is therefore a mechanism of transition from symbolic representation to multimodal feeling.

In this sense, MEM addresses a question left open by RPT. RPT links phenomenal vision to local recurrence and the integrated organization of the percept. MEM adds that recurrence generates phenomenality when it reconstructs receptor-grounded content in lower-level maps and connects that content with bodily state. Not every feedback loop therefore needs to be conscious. Purely regulatory recurrence that does not produce a dominant reconstruction of sensory and interoceptive maps may remain unconscious. This may address the criticism that feedback also occurs without consciousness.

### 7.2. "To Feel, a System Must Have Sensors": The Claim and Its Scientific Specification

Organisms that maintain their homeostatic organization continuously exchange matter and energy with their environment. They must monitor both external changes and their own internal state. To feel anything, a system needs sensors. Exteroceptors provide information about the world, while interoceptors and proprioceptors provide information about the body. A sensor transduces a particular form of energy or physiological change into a neural signal. The nervous system then processes **this** signal by selecting, integrating, and interpreting incoming information in relation to the organism's current state, its history, and regulatory needs.

For an organism to consciously feel a stimulus, sensory information must first be registered and processed by the relevant sensory systems and subsequently integrated with the organism's current homeostatic state. If the stimulus is matters for the organism's needs and homeostatic regulation, its representation can become associated with an affective state and give rise to conscious feeling. In a simple form, this may produce elementary emotional states associated with the stimulus's beneficial or detrimental effect on the organism (Keramati & Gutkin 2014).

In this account, feeling is not a property of arbitrary computation. It is a dynamic conscious state of an organism arising from the interpretation of a stimulus through an appropriately grounded sensory or interoceptive representation. Feeling need not be the final stage of stimulus processing; as an active representational state, it may participate in reactivating representations associated with the modality from which it was formed. This claim requires precision. In standard neurobiology, a receptor does not itself "feel"; it transduces energy or physiological state. Receptor activation alone is also insufficient for phenomenal feeling, as shown by unconscious sensory processing.

MEM therefore requires receptor grounding, appropriate recurrent network dynamics, and a dominant reconstruction. During imagery, the peripheral receptors need not be currently activated. Feedback reactivation of central maps learned through their previous activity is sufficient. The refined principle is therefore that phenomenal content in MEM must be sensory-grounded but need not always be peripherally elicited. Further processes are required, including the selection of competing representations, attentional mechanisms, integration with the organism's current state and homeostatic needs, and recurrent processing leading to the stabilization and reconstruction of the dominant representation. .

This account avoids two extremes. MEM does not reduce feeling to an external stimulus, because dreams, memories, and imagery can be generated internally. Nor does it identify phenomenality with an abstract higher-level representation that lacks a connection to the organism's sensory and bodily groundingSensory grounding is thus a realization condition for phenomenal content in MEM but is not sufficient by itself. The additional realization condition concernsthe dynamics of selection, integration, recurrence, and representational stabilization in the context of the organism's current homeostatic state.

### 7.3. Is Secondary Perception the Only Element Missing from RPT?

The answer is no. Secondary perception is an important missing mechanism from the perspective of phenomenality, but the transition from RPT to MEM requires at least six further elements: a memory-based source of reconstruction; bottom-up/top-down competition; interoceptive and allostatic valuation; coupling with effectors and learning from the consequences of action; focused attention; and multi-cycle dynamics that support deliberation and the construction of a model of reality. The phrase "feedback activates V1" does not yet explain why one pattern rather than another is activated, why it has a particular valence, or how it influences behaviour.

MEM is therefore structurally similar to RPT but differs in its explanatory scope. RPT identifies the type of dynamics required for conscious vision. MEM specifies the possible content, grounding, and function of those dynamics in an embodied system. The relationship is one of partial inclusion: the FFS/RP core of RPT is a component of MEM, but it does not exhaust MEM.

**8. Interoception, Allostasis, and the Perceptual Theory of Emotions**

RPT focuses primarily on vision, whereas MEM treats emotions as a necessary dimension of an autonomous mind. Interoceptors and complex sensory systems provide information about temperature, hunger, thirst, pain, visceral tension, blood gases, pressure, heart rate, and many other states. This information is integrated in the brainstem, hypothalamus, thalamus, insula, cingulate cortex, and limbic structures. Contemporary neuroscience links these pathways to subjective bodily state and homeostatic regulation (Craig, 2002, 2009; Critchley & Garfinkel, 2017; Damasio & Damasio, 2022).

MEM belongs to the family of perceptual theories of emotion. Historical points of reference are James (1884) and Lange (1922/1885) who closely associated emotional feeling with the perception of bodily changes. More recent accounts - for example, Prinz (2004) - understand emotion as the perception, through bodily state, of an evaluatively significant organism–world relationship. Brady (2013) emphasizes the epistemic role of emotional experience, while Salmela (2011) examines the possibilities and limits of modelling emotion on perception. MEM adds a receptor-based and recurrent mechanism: internal states are recognized, associated with percepts, and can be reinstated through secondary perception. During secondary perception, they are associated with a broader range of interoceptive states, which are identical to affective states in perceptual theories of emotion.

Homeostasis concerns the maintenance of variables within biologically viable limits; allostasis concerns the anticipatory and contextual regulation of these limits. Deviations beyond the tolerance ranges generate regulatory pressure. In MEM, global affect is not a separate “mental substance” but a distributed control signal that changes attentional gain, learning rate, memory consolidation, representational valence, and the balance between exploration and exploitation. Consequently, the same object may acquire a different representational strength during hunger, satiety, anxiety, fatigue, or pain.

Emotion is also linked to memory. Co-occurring exteroceptive and interoceptive activations form associated components of a semblion. When the object representation is subsequently reactivated, it can also reconstruct lower-level bodily maps. Feeling during recollection is therefore not a description such as “I was afraid”, but a partial reinstatement of a receptor-grounded fear pattern—that is, a set of interoceptively grounded internal states and regulatory variables. Conversely, a current hormonal or metabolic state changes the matching context and may shift attention towards particular memories, interpretations, and actions.

MEM also invokes an integrated orienting, affective, and motor node, described as a triangle comprising the superior colliculus (SC), periaqueductal grey (PAG), and midbrain locomotor region (MLR). The SC integrates sensory orienting, the PAG participates in organizing defensive responses and affective states, and the MLR links selection with movement. This is a useful functional schema, not evidence that the three structures constitute a single anatomically closed “decision module”. Research on the subcortical foundations of affect and action supports its components (Merker, 2007; Panksepp, 1998; Solms, 2021; Galus, 2024a).

Compared with RPT, this is a qualitative extension. Emotions are not merely additional information integrated with a perceived face. They prioritize and guide learning, influencing what is recognized, how strongly it competes, what is remembered, and which action follows. Embodiment in MEM is therefore not a decorative phenomenology but a condition for organizing the entire processing cycle.

**9. From Access to Action: Effectors, Selection, and Deliberation**

Activations compete not only to determine perceptual content but also for access to effectors. In higher cognitive and motor layers, a winning coalition may activate a learned response program stored in procedural memory. When a configuration is clear, urgent, and strongly valued, transmission proceeds rapidly from receptors through recognition to motor areas. Such action may be reflexive, instinctive, or learned and does not require a phenomenal report. In MEM terminology, it is sometimes associated with the executive aspect of the cognitive system, although the process itself can be functionally unconscious.

If access to motor execution is temporarily inhibited by conflict, attention directed towards internal content, or the absence of a sufficiently effective program, activation may circulate through successive cycles. Each cycle can activate a modified representation of a possible state, reconstruct associated feelings, predict consequences, and again subject alternatives to competition. Deliberation is therefore a dynamic simulation rather than a one-time decision made by a central homunculus.

This allows a cautious comparison between MEM and dual-process theories. The rapid FFS and direct transmission to action resemble automatic System 1, whereas repeated recurrent cycles, response inhibition, and scenario comparison resemble reflective System 2 (Kahneman, 2011; Evans & Stanovich, 2013). These distinctions are not identical. As Galus (2026, footnote 11) notes, System 2 is not simply a single RP pass; it requires repeated iterations extended over time. At the same time, System 1 may also use local feedback loops, and some deliberation remains automatic.

Access consciousness in MEM is largely the result of such functional selection. Content becomes accessible when it dominates working layers and can be associated with language, memory, planning, and movement. Phenomenal consciousness has a different condition: the secondary activation of lower-level modality-specific sensory fields. The two aspects usually cooperate because the same winning representation sends signals both for sensory reconstruction and to executive circuits. They can nevertheless partially dissociate: action may precede reporting, and a vivid experience may not lead to an overt response.

The model preserves causal closure in the physical world. A phenomenal "feeling" need not act as an immaterial cause that activate a neuron. The same processes that are experienced from the first-person perspective as feelings are, from the third-person perspective, map activation, gain changes, and transmission to effectors. In this sense, phenomenality can be locally — in a spatiotemporal sense — an epiphenomenal as a description, while the entire process realizing it has causal power (Galus, 2024a).

## 10. Dynamic Semblions, Image Schemas, and Logical Processes

### 10.1. Representing Relations Extended in Time

A static semblion represents a relatively enduring configuration of an object or concept. A dynamic semblion encodes a sequence of changes: a movement trajectory, causal succession, procedure, rhythm, approach, entry into a container, increasing force, or recovery of balance. MEM postulates a transformation $T$ that converts the temporal intervals between observed events into spatial relations among successive traces in network layers, and an inverse transformation $T^{-1}$ that converts the stored spatial arrangement into a sequence of activations during replay (Galus, 2026, footnote 13). This is an attempt to explain how an episode can be remembered and simulated.

Image schemas—CONTAINER, SOURCE–PATH–GOAL, FORCE, BALANCE, UP–DOWN, and PART–WHOLE—originate primarily from the embodied semantics tradition developed by Mark Johnson and George Lakoff (Johnson, 1987; Lakoff & Johnson, 1980, 1999). In MEM, representations of relations meet representations of emotions because these dynamic schemas are learned in an acting body and, during simulation, can reconstruct both modality-specific images and interoceptive states associated with movement.

Operating on relations requires more than object classification. To understand that $A$ entered $B$, a system must represent the prior location, the change in trajectory, the boundary, and the final state. To predict the consequences of force, it must encode direction, resistance, and outcome. Dynamic semblions are proposed as the neural substrate of such structures. Analogous patterns can subsequently be applied to abstract domains: "entering a problem", "the path to a goal", or "the balance of arguments". Similarity and association provides a basis for metaphor, analogy, and inference (Gentner, 1983; Gentner & Holyoak, 1997).

The claim that logical processes emerge from the recognition of dynamic structures is an important MEM hypothesis. Deduction need not always involve the manipulation of disembodied symbols; it may involve activating a learned transition schema, checking whether its conditions are satisfied, and simulating the consequence. However, formal logic has normative properties that similarity alone does not guarantee. Future psychological research should therefore demonstrate how schema operators preserve truth relations and when the associative mechanism produces errors.

### 10.2. A Model of Reality and the Stream of Thought

Multilayer, compressed, associative representations form a model of known reality. This is not a complete copy of the world but a set of predictable structures: objects, their features, relations, possible changes, significance for the body, and response programs. Perception updates the model, while recurrence and secondary perception use it to complete the current scene. Memory and imagery activate the model without strong external stimulus.

A stream of thought arises when an active semblion activates associated structures laterally or top-down. The next winner reconstructively evokes images, words, feelings, or action schemas, and their activation becomes input to the next cycle. Sometimes the chain maintains a topic through strong semantic relations; at other times, a mental saccade redirects it towards a remote association (Starzyk, 2011). Awareness of this stream is not a separate observer but secondary perception of successive activations.

This description connects access and phenomenal consciousness. The access aspect of the stream consists in automatic transitions between representations and programs. It's phenomenal aspect consists in their modality-specific visualization and feeling through feedback activation of receptor-related fields. RPT explains why recurrence is need for integrated percept but does not develop a comparable account of a sustained internal sequence, dynamic schemas, and logical simulation.

## 11. The Mathematical Model of MEM

Mathematical formalism enables a more precise description of processes and relations between brain structures. For MEM, this role is fulfilled by the article "A Mathematical Model of Motivated Emotional Mind – Cognitive Embodied System", which formalizes MEM at a mesoscopic level. It is not a ready-made algorithm that generates consciousness but a "formal skeleton" of the proposed dependencies and interactions. It specifies classes of variables and relations linking receptor-mediated perception, re-entry, memory, allostatic regulation, and action (Galus & Starzyk, 2026). Formalization does not does not prove that recurrence is identical to consciousness. It can, however, test whether MEM's postulates are mutually consistent, whether RPT can be obtained as a restricted case of MEM dynamics, and what additional falsifiable dependencies MEM introduces.[1]

### 11.1. Layered Dynamics and the Shared RPT–MEM Core

The model describes the state of unit $i$ in layer $\ell$ as depending on five distinct streams: feedforward, lateral, feedback, raw interoception/proprioception $y(t)$, and integrated bodily, motivational, and memory context $b(t)$:

---

[1] Note. This article retains the equation numbers used in the cited mathematical manuscript. The omitted intermediate equations can be found in 'A Mathematical Model of Motivated Emotional Mind – Cognitive Embodied System'. Equation (1R) is the only equation does not present in the mathematical model.

$$h_i^{(\ell)}(t+\Delta t) = f_i^{(\ell)}\left(A_i^{ff,\ell}(t) + A_i^{lat,\ell}(t) + A_i^{fb,\ell}(t) + A_i^{int,\ell}(t) + A_i^{ctx,\ell}(t) + \theta_i^{(\ell)}\right) \tag{1}$$

The terms $A^{\mathrm{ff}}, A^{\mathrm{lat}}, A^{\mathrm{fb}}, A^{\mathrm{int}}$ $i$ $A^{\mathrm{ctx}}$ are weighted contributions of the corresponding streams. The symbol $\theta_i^{(\ell)}$ should be understood here as a bias of either sign—a baseline offset. This distinction removes the ambiguity concerning its sign.[2]

The core consistent with RPT is obtained by disabling the contributions specific to embodied MEM, $A^{\mathrm{int}} = A^{\mathrm{ctx}} = 0$:

$$h_i^{(\ell)}(t+\Delta t) = f_i^{(\ell)}\left(A_i^{ff,\ell}(t) + A_i^{lat,\ell}(t) + A_i^{fb,\ell}(t) + \theta_i^{(\ell)}\right) \tag{1R}$$

The RPT-compatible core is obtained by disabling contributions specific to embodied MEM.  During the initial FFS, the contribution $A^{\mathrm{fb}}$ is absent or too weak to alter the state of the lower layers. Once feedback signal $A^{\mathrm{fb}} \neq 0$ arrives, the system enters recurrent processing. The restricted form of equation (1R) does not prove all of RPT, but shows how its minimal dynamic core: - feedforward transmission, lateral interactions, and feedback - can be included in MEM. The fullmodel adds interoceptive and contextual streams.

### 11.2. Allostasis and Affect

The manuscript's notation distinguishes raw bodily signals $y(t)$, regulated allostatic variables $n(t)$, and integrated context $b(t)$. For the lower and upper tolerance limits $n_{\mathrm{r}}^{\mathrm{min}}(t), n_{\mathrm{r}}^{\mathrm{max}}(t)$, the one-sided violations are:

$$\Delta n_r^-(t) = \max\left(0, n_r^{\min}(t) - n_r(t)\right), \qquad \Delta n_r^+(t) = \max\left(0, n_r(t) - n_r^{\max}(t)\right) \tag{2}$$

On this basis, we can determine the vector of allostatic violations, affect, and the cost to the system:

$$d(t) = \left(\Delta n_1^-(t), \dots, \Delta n_R^-(t), \Delta n_1^+(t), \dots, \Delta n_R^+(t)\right)^{\top}, \qquad A(t) = g\left(d(t)\right) = \tanh(w_A^{\top} d(t) + c_A) \tag{3}$$

$$C(t) = \sum_{r=1}^{R}\left[\alpha_r\left(\Delta n_r^-(t)\right)^2 + \beta_r\left(\Delta n_r^+(t)\right)^2\right] \tag{4}$$

The vector $d(t)$ collects all violations, while global affect $A(t)$ constitutes a bounded regulatory signal. The cost $C(t)$ remains non-negative if $\alpha_{\mathrm{r}}, \beta_{\mathrm{r}} \geq 0$:

Tolerance limits may depend on time and context, so the model does not assume a single fixed equilibrium point. $A(t)$ is not the name of a specific emotion but a signal of regulatory pressure: its magnitude influences, among other things, the learning rate and exploration, while its sign can update the valence of active semblions. This link between bodily regulation, plasticity, and representational selection extends beyond the scope of RPT.

[2] Variable notation follows the cited article.

### 11.3. Semblion Matching, Competition, and Selection

A semblion $S_{\mathrm{m}} = (\mu_{\mathrm{m}}, \rho_{\mathrm{m}}, v_{\mathrm{m}})$ links a sensory prototype $\mu_{\mathrm{m}}$, an established bodily and motivational context $\rho_{\mathrm{m}}$, and a valence $v_{\mathrm{m}}$. Its matching score retains exactly the notation of Equation (6):

$$u_m(t) = \alpha_s \,\mathrm{sim}\big(h_{sens}^{ff}(t), \mu_m\big) + \beta_b \,\mathrm{sim}\big(b(t), \rho_m\big) + \xi_v v_m \tag{6}$$

Its strength is jointly determined by receptor evidence, the match between the current context $b(t)$ and $\rho_{\mathrm{m}}$, and the Equation (6) is logically important: the representation is not determined solely by the current stimulus. established value $v_m$. TopK in (8) creates only a set of candidates $\mathcal{C}(t)$, rather than a set of final winners:

$$\mathcal{C}(t) = \mathrm{TopK}_K(\{u_m(t)\}_{m=1}^{M}) \tag{8}$$

A variant of soft competition is then applied within the candidate set $\mathcal{C}(t)$. Lateral associative interactions $L_{\mathrm{mn}}(t)$ modify the raw matching score before normalization:

(The three components of the original Equation (9) are written as (9a) – (9c).)

$$q_m(t) = \sum_{\substack{n\in\mathcal{C}(t)\\ n\neq m}} L_{mn}(t)[u_n(t)]_+ \tag{9a}$$

$$\tilde{u}_m(t) = u_m(t) + \lambda_L q_m(t) \tag{9b}$$

$$a_m(t) = \begin{cases} \dfrac{\exp(\tilde{u}_m(t)/\tau_c)}{\sum_{j\in\mathcal{C}(t)} \exp\big(\tilde{u}_j(t)/\tau_c\big)}, & m \in \mathcal{C}(t), \\ 0, & m \notin \mathcal{C}(t). \end{cases} \tag{9c}$$

where $[u_n(t)]_+ = max(0, u_n(t))$, $\lambda_L$ is the strength of lateral priming, and $\tau_c > 0$ is the competition temperature. Only the dominance threshold $\Theta$ defines the set $W(t)$ capable of driving top-down reconstruction:

$$W(t) = \{m: a_m(t) > \Theta\} \tag{12}$$

The dominance threshold $\Theta$ [3] is positive and less than one ($0 < \Theta < 1$), because it refers to the activation shares $a_m(t)$ determined by softmax. Depending on the activation distribution and values, the set $W(t)$ may contain one semblion, several semblions, or remain empty.

Distinguishing $\mathcal{C}(t)$ from $W(t)$ prevents preliminary candidate selection from being confused with final winning coalition. WTA is a limiting case of sharp competition; the model also allows several compatible semblions to co-dominate.

[3] This symbol should be distinguished from the baseline offset $\theta_i^{(\ell)}$ in Equation (1), which may have either sign.

### 11.4. Reconstruction, Re-Entry, and Secondary Perception

Winning semblions reconstruct the modality-specific prototype, while the effective representation combines the feedforward signal with the reconstruction:

$$\hat{h}_{sens}(t) = \sum_{m \in W(t)} a_m(t)\mu_m \tag{13}$$

$$h_{sens}^{eff}(t) = (1-\lambda)h_{sens}^{ff}(t) + \lambda\hat{h}_{sens}(t), \qquad 0 \le \lambda \le 1 \tag{14}$$

In Equation (14), $\lambda$ is a constant signal-integration parameter. A change in the dominance of either signal may result from a change in the amplitude of $h_{sens}^{ff}(t)$ *or* $\hat{h}_{sens}(t)$. Different values of $\lambda$ may describe different fixed regimes of the model.

Equations (13) – (14) formalize reconstruction, but the manuscript itself does not provide an explicit criterion for the transition from reconstruction to secondary perception. In a stronger formulation of MEM, the following operational condition can be introduced:

$$\chi_{SP}(t) = 1 \quad \left[W(t) \neq \varnothing \ \wedge \parallel \hat{h}_{sens}(t) \parallel \geq \Theta_{rec} \ \wedge \parallel h_{int}^{act}(t) \parallel \geq \Theta_{int}\right]; \ \text{ otherwise } \ 0$$

The variable $\chi_{SP}(t) = 1$ indicates the occurrence of secondary perception; $h_{int}^{act}(t)$ denotes the current activation of modality-specific interoceptive maps, and $\Theta_{rec}\ i\ \Theta_{int}$ are empirically determined thresholds for reconstruction strength and interoceptive co-activation, respectively.

A stronger formulation of MEM can introduce an operational condition: secondary perception occurs when reconstruction strength and co-activation of modal and interoceptive maps exceed empirically determined thresholds.. This condition is a testable hypothesis, not an algebraic consequence of the preceding equations. It identifies re-entry together with sufficient sensory and interoceptive activation as a proposed condition of phenomenal experience. This is precisely where MEM goes beyond RPT, which assigns recurrence a fundamental role in conscious vision but does not formalize the memory-based source of reconstruction, secondary perception, or its affective grounding.

### 11.5. Regulatory Cost, Value, and the Procedural Gap

A program $\Pi$ is a sequence or composition of elementary transition operators, and the decision state takes the form $s_t = (h(t), b(t), n(t))$. The predicted cost of the program over the horizon $T$ is specified by Equation (21a):

$$J(\Pi \mid s_t) = \mathbb{E}\left[\sum_{\tau=0}^{T} \gamma^{\tau}\, C(t+\tau\Delta t; \Pi) \mid s_t\right] \tag{21a}$$

Since $C(t) \geq 0$, we have $J(\Pi|s_t) \geq 0$. The decision value $Q(\Pi \mid s_t)$ is defined as negative cost: the lower the cost, the higher the decision value.

A procedural gap $K(t) = 1$ occurs when no known program reaches the sufficiency threshold $\theta_Q$:

$$K(t) = 1 \quad \Leftrightarrow \quad \min_{\Pi \in \mathcal{P}(t)} J(\Pi \mid s_t) > \theta_Q \tag{28}$$

For a non-trivial sufficiency threshold, we must assume $\theta_Q > 0$.

The set of known programs $\mathcal{P}(t)$, the policy distribution $\pi(\Pi|s_t)$, and the set of newly generated candidates $\mathcal{P}_{cand(t)}$ are distinct objects. The best candidate is equivalently selected by minimizing the relevant expected cost $J$:

$$\Pi_{new}(t) = \underset{\Pi \in \mathcal{P}_{cand}(t)}{\arg\min} J(\Pi \mid s_t) \tag{29}$$

Deliberation across multiple cycles corresponds to the repeated execution of dynamics (1), reconstruction (13) – (14), and program simulation; it is not a single recurrent processing pass.

### 11.6. Implications of the Formalization: The Preserved Core and Extensions Beyond RPT

The formalization supports a relationship of partial inclusion not theoretical identity. The mathematical core shared with RPT is preserved in Equation (1R): the FFS provides rapid bottom-up activation, lateral interactions support local organization, and feedback introduces recurrence and modifies the lower layers. Disabling $A^{\text{int}}\, i\, A^{\text{ctx}}$ reduces MEM to dynamics with a scope close to that of RPT, without removing feedforward transmission or re-entry.

MEM extends beyond RPT in five formally identifiable respects: (1) it distinguishes raw interoception $y(t)$, the allostatic state $n(t)$, and integrated context $b(t)$; (2) links recognition to a sensory prototype, context, and valence in Equation (6); (3) separates candidates $C(t)$, associative competition, and winners $W(t)$ that initiate reconstruction; (4) introduces memory-based reconstruction and an operational condition for secondary perception with an interoceptive component; and (5) connects successive cycles to regulatory cost, exploration, a procedural gap, simulation, and action selection.

The formalism yields distinguishing predictions. Disrupting feedback while preserving the FFS should impair reconstruction and phenomenal reporting more strongly than rapid unconscious detection. Changing allostatic state should, after controlling for the external stimulus, alter semblion matching through memory context, valence, learning rate, vividness of reconstruction, exploration, and program selection. Distinguishing $C(t)$ from $W(t)$, in turn, predicts cases in which several representations are initially active but only a subset crosses the reconstruction threshold. Such dependencies are not derived from the minimal dynamics of RPT alone.

The strongest justified conclusion is therefore that the mathematical model demonstrates the architectural coherence of MEM and shows how to preserve the recurrent core of RPT within a broader perceptual, allostatic, and decision-making system. It does not mathematically prove the identity of secondary perception and phenomenal consciousness. Instead, it turns claim into an explicit operational condition that can be tested through perturbation, neuroimaging, and simulation studies.

## 12. A Systematic Comparison of RPT and MEM

Table 1 summarizes the relationship between the two theories, distinguishing their shared core from differences in explanatory level.

**Table 1. RPT and MEM — The Shared Core and Differences in the Level of Explanation**

| Dimension | RPT | MEM | Relationship |
|---|---|---|---|
| Scope | Conscious vision and the transition from unconscious to conscious processing. | The embodied mind: perception, emotions, memory, thinking, planning, and action. | MEM extends the domain, rather than merely a parameter of RPT. |
| FFS | Rapid feature and category extraction can be unconscious. | Bottom-up processing: recognition in semblions, competition, and potentially rapid action. | Strong agreement at the core. |
| Recurrence | Figure–ground organization, integration, and stabilization of the conscious percept. | Memory-based reconstruction of lower-level maps, perceptual correction, feelings, simulation, and deliberation. | MEM specifies the content and functions of feedback. |
| Phenomenality | Local RP is a leading candidate mechanism of P-consciousness. | Dominant secondary perception of sensory and interoceptive maps constitutes experience. | MEM formulates a stronger realization condition. |
| Attention | Distinct from consciousness; influences access and reporting. | Emerges from competition involving signal strength, matching, affect, needs, and context. | MEM adds a mechanism of prioritization. |
| Memory | Learning and plasticity are included (LEARN), but no memory architecture is proposed. | Semblions store static and dynamic engrams; memory corrects perception. | A structural extension. |
| The body | Bodily states may be integrated; they are not the organizing principle of the theory. | Receptors, allostasis, hormones, autonomic processes, and effectors are constitutive. | An ontological and biological extension. |
| Emotions | Emotional content can be recurrently integrated with the percept. | Emotions are the perception and valuation of bodily state; they modulate learning and selection. | MEM assigns emotions a regulatory role. |
| The external world versus thought | Competition among percepts and the influence of attention. | Explicit competition between bottom-up and top-down signals; a perception–imagery–hallucination continuum. | A new mechanistic variable. |

| Dimension | RPT | MEM | Relationship |
|---|---|---|---|
| Action | Not the principal explanandum. | Competition for effectors; a rapid response or its inhibition and the simulation of alternatives. | A sensorimotor extension. |
| System 1/2 | No explicit central mapping. | FFS resembles System 1; deliberation across multiple cycles resembles System 2. | Analogous, but not identical. |
| Logic | Not addressed by the theory. | Dynamic semblions and image schemas as operators of relations and consequences. | A cognitive extension. |
| Mathematical formalization | Temporal and network models of recurrence; an empirical theory of visual cortex. | A formal skeleton of FF/lateral/FB processing, allostasis, competition, reconstruction, and policy. | MEM broadens the formal scope but requires simulations. |
| Evidential status | An extensive experimental literature on vision; debate over the necessity and sufficiency of RP. | Components consistent with neurobiology; semblion integration and the sufficiency of secondary perception remain hypothetical. | RPT is empirically more mature within its domain. |

Sources: Lamme and Roelfsema (2000), Lamme (2015, 2020), Galus and Starzyk (2021), Galus (2018, 2024a, 2024b, 2025a, 2025b, 2026).

**13. How MEM Complements RPT**

The first level is descriptive. MEM covers phenomena beyond the principal scope of RPT: memory, imagery, dreams, hallucinations, emotions, motivation, attention, planning, and action. Merely extending the list of phenomena would not constitute a substantive advantage. MEM's significance depends on whether it identify a common mechanism generating these functions.

The second level is architectural. Semblions connect receptor, representational, emotional, and motor layers. Convergence compresses information, associations construct a model of the world, and feedback connections reconstruct modality-specific content. These functions are not assigned to separate boxes; they are expected to emerge from the propagation of activation within a single structure. The aim is to explain the mechanisms that implement attention or feeling, rather than simply naming these functions.

The third level is biological. In MEM, decisions are constrained by bodily state, allostatic thresholds, interoceptive pathways, plasticity, and effectors. Established mechanisms— including LTP, neurogenesis and synaptogenesis, dendritic spine remodelling, lateral inhibition, neuromodulation, sensory maps, cortical feedback, and subcortical circuits—provide possible components of an implementation. Their specific organization of these mechanisms into semblions remains a hypothesis, but the model identifies processes that could generate selection and learning.

The fourth level is phenomenological. RPT links consciousness to recurrent organization. MEM proposes that qualitative content arises from the secondary activation of modality-specific maps and associations with interoceptive representations. A memory thus has an appearance, and an emotion has a feeling, because sensory and interoceptive patterns are reactivated rather than merely represented by abstract labels. This is the most important and also the riskiest step.

The fifth level is dynamic and pragmatic. Content competes for further transmission, attention, and effectors. A strong external stimulus can suppress thoughts; a strong reconstruction can block access to the current scene; and conflict can inhibit a response and initiate repeated simulations. Consequently, the same mechanism links attention switching, the stream of thought, decisions, and behaviour.

The sixth level is formal. The mathematical model of MEM expresses competition, reconstruction, allostasis, affect, and action policy as variables and operators. This opens the way to simulations, stability analysis, and model comparisons. It is not, however, proof of the theory's truth: formalization can demonstrate coherence and generate predictions, but further research must connect it with data.

**14. Discriminating Predictions and a Research Program**

For MEM to be more than a broad narrative, it should predict outcomes different from those of minimal RPT. First, selectively disrupting feedback to early sensory maps should impair the vividness and modality-specific quality of imagery, memories, and dreams more strongly than the effectiveness of rapid unconscious recognition. If phenomenal imagery remains unchanged despite effective disruption of modal reconstruction, the central MEM claim would be weakened.

Second, interoceptive and allostatic manipulations - while controlling for external reward and task demands—should alter representational dominance, the valence of associations, the vividness of affective recollection, and exploration strategy. MEM predicts not merely a correlation between bodily state and reporting, but a systematic influence on competition and learning.

Third, the strength of external activation should exhibit measurable, nonlinear competition with top-down content. Under intense modality-specific stimulation, activity and reports associated with subtle imagery should decline. When internal attention is sustained, access to irrelevant stimuli should decline despite the preservation of early FFS. The fusion parameter $\lambda$ in the mathematical model can be fitted to neural and behavioural data.

Fourth, multi-cycle deliberation should involve alternating reconstructions of possible outcomes and their corresponding bodily states before selection. Blocking the motor response without blocking reconstruction should prolong simulation, whereas requiring a rapid response should favour the FFS pathway. This is more precise than simply saying that System 2 is slow.

Fifth, the theory of semblions requires the identification of distributed but reproducible coalitions spanning modality-specific, semantic, affective, and motor layers. Multi-area experiments and closed-loop perturbations should test whether activation of a higher-level component predictably reconstructs lower-level components and whether shared learning history creates the proposed structure. fMRI correlations alone would not be sufficient.

Sixth, the model should specify boundary cases. If local recurrence occurs without receptor-grounded reconstruction, MEM may predict the absence of phenomenality despite satisfaction of the minimal RPT criterion. If reconstruction is strong but does not lead to global reporting, MEM predicts possible phenomenality without access. Such dissociations enable empirical comparisons between theories rather than defining outcomes after the fact.

**Table 2. The Epistemic Status of the Principal Elements of the Argument**

| Level | Assessment |
|---|---|
| Well-established components | Hierarchical and feedback cortical connections; FFS; figure–ground modulation; synaptic plasticity; interoception; top-down sensory reactivation; competition among representations. |
| Integrative MEM hypotheses | Semblions as the appropriate unit of representation; a shared sensory, semantic, affective, and motor architecture; dynamic semblions and $T/T^{-1}$ transformations. |
| Strong constitutive claims | Secondary perception as the basis of phenomenality; emotions as a receptor-mediated perceptual reading of allostasis; sensory grounding as a condition of qualia. |
| Formalization | A coherent mesoscopic skeleton of variables and operators; a full implementation, parameter estimation, and comparative validation are still lacking. |
| Theoretical risks | An overly broad explanandum; the possibility of post hoc fitting; an ambiguous microphysical implementation of semblions; difficulty measuring phenomenality independently. |

### 15. Limitations of Both Theories

RPT is limited in its scope and its criterion of sufficiency. It explains important aspects of visual-system functioning, but the term "recurrent processing" encompasses many loops with different functions. The theory should specify which types, layers, timings, and directions of recurrence are constitutive. Separating phenomenality from reporting protects the theory against equating consciousness with behaviour, but makes independent verification more difficult.

MEM faces the opposite problem: its scope is very broad. The model encompasses perception, memory, emotions, motivation, attention, creativity, and action. This integration is its strength, but increases the risk that every phenomenon will be described through a flexible combination of activations. Mathematical formalization should constrain this flexibility by specifying parameters, training data, and conditions of falsification.

Not all claims concerning sensors are unambiguous. Pain, moods, and diffuse affective states cannot always be attributed to a single receptor or topographic map. The concept of a sensory complex is functionally useful but may become tautological if every feeling-capable structure is defined as a sensor. Independent criteria are required: transduction, specific pathways, central maps, and causal influence on experience.

The proposed neuronal–astrocytic structure of semblions is promising, but the literature on tripartite synapses and plasticity does not yet demonstrate the representational architecture of MEM. Possible substrates must be distinguished from confirmed ones. Similarly, concept cells support sparse and abstract coding but do not establish the precise semi-hierarchy.

Finally, even if secondary perception predicts experiential vividness perfectly, the philosophical question remains whether it explains why map activation is felt or merely identifies its most immediate correlate? MEM offers a reductive answer: experience is the process viewed from the first-person perspective. Critics may regard this as an identity hypothesis requiring further justification rather than a logical solution to Chalmers's hard problem (Chalmers, 1995). The scientific value of the model lies in translating this hypothesis into mechanisms amenable to perturbation.

## 16. Conclusion

RPT remains a fundamental point of reference because it has shown that rapid feedforward processing and recognition are not equivalent to conscious vision, and that the recurrent organization of representations is crucial for the transition towards a percept. Its contribution is precise, experimentally fruitful, and particularly strong in the visual domain.

MEM accepts this core but develops it into an embodied architecture. Lower-level fields are activated by exteroception and interoception. Higher layers recognize configurations and link them to memory, hormonal and metabolic state, emotions, and action programs. Activations compete for dominance and access to effectors. Feedback reconstructs patterns in lower-level maps. Repeated cycles enable correction, recollection, imagery, and deliberation. Semi-hierarchical semblions compress and associate information, constructing a model of the world and dynamic schemas of relations.

MEM can therefore be regarded as similar to RPT in its description of the FFS and recurrent stimulation, but this similarity encompasses only the temporal skeleton. From the MEM perspective, RPT lacks more than secondary perception. It also lacks a developed receptor-based mechanism, grounding of qualia, allostatic perception of emotions, sources of representational weighting, memory-based correction, sensorimotor competition, dynamic semblions, and multi-cycle deliberation. MEM addresses these gaps through an integrative theory.

The most distinctive MEM claim is that what an organism feels as an image, sound, pain, mood, or visualization of thought is the secondary activation of networks shaped by sensory contact with the world and the body. Exteroceptive grounding gives experiences their external modality; interoceptive grounding gives them bodily and affective valence. Feeling and imagery can occur without a current stimulus through central reconstruction, but without a history of sensory interaction and related maps, content would lack the qualitative anchoring proposed by MEM.

The strength of MEM therefore lies in its attempt to derive mental functions from the dynamics of a specific biological structure rather than merely enumerate them. Its scientific future depends on meeting the empirical demands of this ambition: demonstrating that semblions, secondary perception, and allostatic control are not merely compatible with data but causally necessary, quantitatively modellable, and explanatory relative to alternatives. Only then can MEM's relationship to RPT be assessed not merely as broad extension, but as an explanation of phenomenal consciousness.